\documentclass[]{aastex631}

\usepackage[T1]{fontenc}
\usepackage{soul}
\usepackage{mathtools}

\shorttitle{True Polar Wander of Lava Planets}
\shortauthors{Kang et al.}
\graphicspath{{./}{Figures/}}

\begin{document}

\title{True Polar Wander of Lava Worlds.}

\correspondingauthor{Wanying Kang}
\email{wanying@mit.edu}

\author[0000-0002-4615-3702]{Wanying Kang}
\affiliation{Earth, Atmospheric and Planetary Science Department, 
  Massachusetts Institute of Technology,
  Cambridge, MA 02139, USA}

\author{Francis Nimmo}
\affiliation{Earth \& Planetary Sciences Department,
  University of California Santa Cruz,
  Santa Cruz, CA 95064, USA}

\author{Feng Ding}
\affiliation{Dept. of Atmospheric and Oceanic Sciences, School of Physics,
  Peking University,
  Beijing 100871, People’s Republic of China}



\begin{abstract}
  As one of the most detectable types of terrestrial planets, lava worlds are highly prioritized targets for exoplanet atmosphere characterization since their atmospheres may reveal what they are made of and how. Our work examines the possibility of true polar wander (TPW) occurring on these ultra-hot tidally-locked planets, powered by mass redistribution from atmospheric flow between the hot permanent day-side and the cold permanent night-side. We find that lava planets within a certain mass and temperature range may undergo TPW, and this likelihood increases with star mass. As a result of TPW, the magma ocean and atmospheric compositions may be less evolved (refractory-enriched) than previously thought and may be validated by exoplanet demographic surveys.
\end{abstract}



\section{Introduction}

Hundreds of rocky planets with substellar equilibrium temperature beyond the melting point of mantle material have been discovered according to the NASA Exoplanets Archive. Their high surface temperature and strong thermal emission make them easy to detect. Especially, with the high-resolution spectrometer carried by James Webb Space Telescope (JWST), atmospheric characterization becomes ever more accessible \citep{Hu-Brandeker-Damiano-et-al-2021:determining, Kreidberg-Hu-Kite-et-al-2021:search, Brandeker-Alibert-Bourrier-et-al-2021:is, Dang-Cowan-Hammond-et-al-2021:hell, Zhang-Dai-Hu-et-al-2021:first, Batalha-Teske-Alam-et-al-2021:seeing}. Since atmospheres on lava planets contain vapor from the day-side magma ocean, they may contain useful information about the composition of their mantle, which is key to understanding planet formation \citep{Wordsworth-Kreidberg-2022:atmospheres}.

Many previous works have studied atmospheric transport on lava planets \citep{Castan-Menou-2011:atmospheres, Kite-Fegley-Schaefer-et-al-2016:atmosphere, Nguyen-Cowan-Banerjee-et-al-2020:modelling, Kang-Ding-Wordsworth-et-al-2021:escaping, Nguyen-Cowan-Pierrehumbert-et-al-2022:impact}. Driven by the pressure gradient between the hot day-side and the cold night-side, mineral gases evaporated from the magma ocean can be accelerated to several thousands of meters per second (supersonic) tens of angular degrees away from the substellar point. This flow will not completely come to a stop at the magma ocean edge because a surface temperature of 1600~K (freezing point of magma ocean) is still beyond the condensation temperature of any single composition in the atmosphere and the mass exchange rate between the atmosphere and the solid surface is limited by the frequency of molecular collision, which is not infinite \citep{Ingersoll-Summers-Schlipf-1985:supersonic, Kang-Ding-Wordsworth-et-al-2021:escaping}. This means that the atmospheric flow can transport a significant amount of mass from the magma ocean and deposit it on the solid lithosphere elsewhere, resulting in a substantial mass redistribution over time. This mass redistribution could potentially alter the shape of the planet and cause it to resemble a flattened donut facing its host star. A donut shape opposes the tendency of the planet's most elongated axis to point in the direction of the star and thus could trigger true polar wander (TPW), in which the solid planet reorients itself relative to the spin axis (which stays fixed in inertial space). The purpose of this work is to quantify the likelihood of TPW on lava planets surrounding different types of stars. We expect the mass of the star to have a significant impact because a more massive star emits far more radiation ($\propto M_\star^3$) and can heat a planet located further away to beyond its melting point. At farther distances, the static tidal bulge, which inhibits TPW, will be much weaker. 

\section{Method and results}

To evaluate the possibility for true polar wander (TPW) to occur on a lava planet, we take the following steps: 1) we obtain the mass deposition rate $F$ by atmospheric flow using the dayside-to-nightside transport model by \citep{Kang-Ding-Wordsworth-et-al-2021:escaping}, 2) we estimate the lithosphere relaxation timescale $\tau_{\mathrm{rlx}}$ for load removal, and multiply $F$ with it to obtain an estimate for the equilibrium mass load $M^{L}$, and 3) we use the TPW criterion derived by \citet{Matsuyama-Nimmo-2007:rotational} and \citet{Nimmo-Matsuyama-2007:reorientation} to determine whether TPW can happen. Instead of focusing on a specific exoplanet, our goal here is to do a general exploration of the parameter space. The parameters we consider here include the insolation received by the planet measured by the substellar point temperature $T_{s0}$, the planetary mass $M_p$ and the mass of the star $M_\star$. As we alter the planetary mass, we also adjust the bulk density accordingly following \citep{Zeng-Sasselov-Jacobsen-2016:mass}. Although $M_\star$ may not directly affect atmospheric mass transport and the resulting mass load $M^L$, it can influence the orbital radius of a planet with a specific $T_{s0}$, and therefore, the static tidal bulge. 

\begin{table*}[hptb!]
  
  \centering
  \begin{tabular}{lll}
  \hline
    Symbol & Name & Definition/Value\\
    \hline
    $M_\star$ & mass of the host star & 1, 3, 10 $\times M_{\mathrm{sun}}$\\
    $R_\star$ & radius of the host star & $R_{\mathrm{sun}}(M_\star/M_{\mathrm{sun}})^{5/6}$ \citep{Phillips-2013:physics}\\
    $L_\star$ & luminosity of the host star & $L_{\mathrm{sun}}(M_\star/M_\mathrm{sun})^3$ \citep{Phillips-2013:physics}\\
    $T_\star$ & illumination temperature of the host star & $\left(L_\star/(4\pi \sigma R_\star^2)\right)^{1/4}$\\
    $\tau_s$ & lifetime of the star & $\tau_{\mathrm{sun}}(M_\star/M_{\mathrm{sun}})^{-2}$\\
    $M_p$ & mass of the planet & 0.01, 0.02, 0.05, 0.1, 0.2, 0.5, 1, 3 $\times M_{e}$\\
    $\rho_p$ & bulk density of the planet & 0.72, 0.73, 0.74, 0.75, 0.815, 0.94, 1.22 $\times\rho_{e}$ \citep{Zeng-Sasselov-Jacobsen-2016:mass}\\
    $a$ & radius of the planet & $(3M_p/(4\pi\rho_p))^{1/3}$\\
    $g$ & gravity of the planet & $GM_p/a^2$ \\
    $\rho_{\mathrm{core}}/\rho_{\mathrm{mantle}}$ & density ratio between the core and the mantle & 2\\
    $a_{\mathrm{core}}$ & core radius & $a/2$\\
    $\mu_{\mathrm{mantle}}$ & mantle rigidity & $2\times 10^{11}$~Pa\\
    $\mu_{\mathrm{core}}$ & core rigidity & $2\times 10^{4}$~Pa \\
    $\eta_{\mathrm{mantle}}$ & mantle viscosity & $10^{21}$~Pa$\cdot$s\\
    $T_{s0}$ & substellar point temperature (radiative equilibrium) & 2000-4000K at 200K increment\\
    $T_s$ & planet surface temperature (radiative equilibrium) & Eq.\ref{eq:Ts}\\
    $d$ & distance between the planet and the star & $\sqrt{L_\star/(4\pi\sigma T_{s0}^4)}+a$\\
    $\Omega$ & orbital/rotation frequency & $\sqrt{GM_\star/d^3}$\\
    $V_a,\ P_a,\ T_a$ & atmospheric velocity, pressure, temperature & output from atmosphere transport model \citep{Kang-Ding-Wordsworth-et-al-2021:escaping}\\
    $F$ & evaporation rate from the magma ocean & Eq.\ref{eq:F-P}\\
    $\alpha$ & mass exchange coefficient between atm \& ocean & 1\\
    $\tilde{F}$& modified mass deposition rate & Fig.\ref{fig:example-case-mass-deposition-load-width}b\\
    $\tau_\mathrm{rlx}$ & relaxation timescale & Eq.\ref{eq:tau-rlx}, Table.~\ref{tab:relaxation-timescale}\\
    $k_2$ & degree-2 love number of the planet & two-layer model formula given by \citet{Harrison-1963:analysis}\\
    $Q_f$ & quality factor of the planet & 100\\
    $\tau_{\mathrm{lock}}$ & timescale for planet to get tidally-locked & Eq.\ref{eq:taulock}\\
    $M^L$ & equilibrium mass load & Eq.\ref{eq:mass-load}\\
    $k_L$ & wavenumber of mass load & Fig.\ref{fig:example-case-mass-deposition-load-width}c\\
    $G_{20}^L$ & non-dimensional $Y_{20}$ mode of the mass load  &  Eq.\ref{eq:G20}\\
    $Q$ & non-dimensional criterion for true polar wander & Eq.\ref{eq:Q-criterion}\\
    \hline
    $M_{\mathrm{sun}}$ & mass of the sun & $2\times 10^{30}$~kg\\
    $R_{\mathrm{sun}}$ & radius of the sun & $7\times 10^{5}$~km\\
    $\tau_{\mathrm{sun}}$ & lifetime of the sun & 10~Ga\\
    $M_e$ & mass of the earth & $6\times 10^{24}$~kg\\
    $\rho_e$ & density of the earth &$5110$~kg/m$^3$\\
    $G$ & gravitational constant & $6.67\times 10^{-11}$ m$^3$ kg$^{-1}$ s$^{-2}$\\
    $\mu$ & molecular weight of SiO & 0.044~g/mol\\
    $R$ & gas constant & $R^*/\mu=189~\mathrm{kg} ~ \mathrm{m}^{2}  \mathrm{s}^{-2}  \mathrm{K}^{-1} \mathrm{kg}^{-1}$\\
    $\sigma$ & Stefan-Boltzman constant & $5.67\times 10^{-8}~W /\left(m^{2} K^{4}\right)$\\
    \hline
     \end{tabular}
  \caption{Parameter definitions. }
  \label{tab:parameters}
  
\end{table*}

\subsection{Atmospheric mass transport.}
Driven by the pressure gradient between the hot day-side and cold night-side, the atmosphere will be rapidly accelerated toward the night-side beyond the sound speed.The governing equations of the atmospheric flow consist of the conservation of vertically-integrated mass flux, momentum flux and mechanical plus dry internal energy flux as derived by \citet{Ingersoll-Summers-Schlipf-1985:supersonic}. Later on, \citet{Kang-Ding-Wordsworth-et-al-2021:escaping} generalized the form to account for condensation and the associated latent heat release, motivated by the fact that magma vapor can be highly condensible. Instead of digging into too many technical details, which can be found in the appendix B of \citet{Kang-Ding-Wordsworth-et-al-2021:escaping}, we will present an example case and discuss how the mass redistribution rate varies with parameters $T_{s0},\ M_p$ and $M_\star$.

\begin{figure*}[htbp!]
    \centering
    \includegraphics[width=\textwidth]{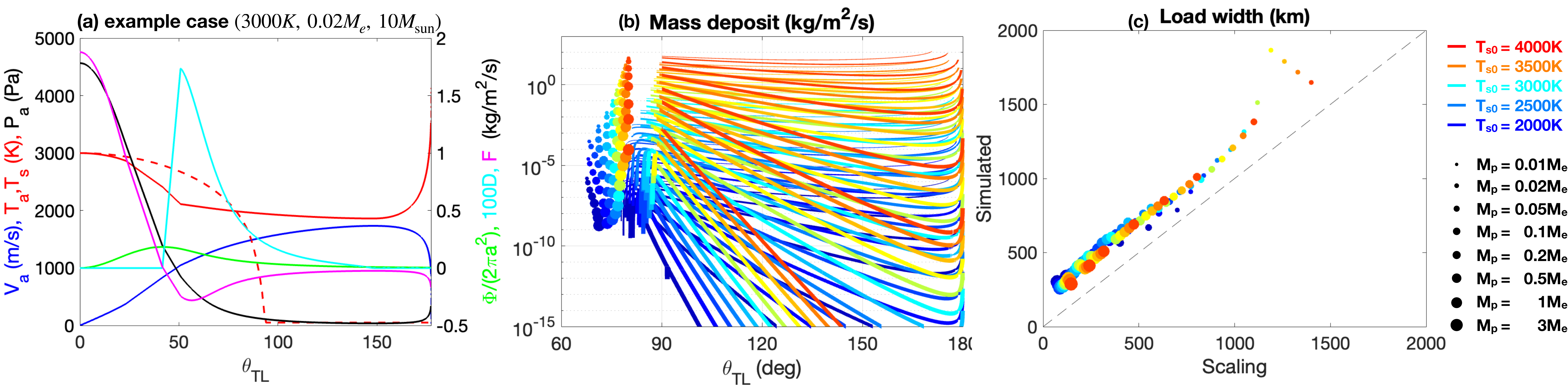}
    \caption{\small{Panel (a) shows the air temperature $T_a$ (red, left), pressure $P_a$ (black, left), velocity $V_a$ (blue, left), surface temperature (red dashed, left), mass transport flux $\Phi$ (green, right), condensation $D$ (cyan, right) and surface mass flux $F$ (magenta, right) for an example case, where $T_{s0}=3000$K, $M_p=0.02~M_e$ and $M_\star=10~M_{\mathrm{sun}}$. Other stellar parameters can be derived from $T_{s0}$ and $M_\star$ following formula given in Table.~\ref{tab:parameters}. Panel (b) shows the mass load profile. Mass deposited to the planetary surface by atmospheric flow out of the magma ocean can be sustained by the lithosphere, as shown by solid curves. Within the magma ocean edge ($T_s>T_f=1600$K), the total mass loss rate per area is assumed to be evenly distributed, as shown by the vertical locations of the solid dots. The horizontal locations are set to the magma ocean edge minus 10~degree. Panel (c) compares the measured e-folding scale of the mass load out of the magma ocean $l^{L}$ against the theoretical prediction $\pi R T_{s0}/g$. For panels (b) and (c), color indicates the temperature at the substellar point of the planet and line width, symbol size reflects the size of the planet; stellar mass $M_\star$ is set to $10M_{\mathrm{sun}}$. }}
    \label{fig:example-case-mass-deposition-load-width}
  \end{figure*}

Fig.~\ref{fig:example-case-mass-deposition-load-width}a shows the air temperature $T_a$, pressure $P_a$, and velocity $V_a$ near the surface as a function of angular distance from the substellar point on the left y-axis and the mass transport $\Phi$, the condensation rate $D$, and most importantly, the surface evaporation/absorption rate $F$ on the right y-axis for an example case, where $T_{s0}=3000$~K and $M_p=0.02$ earth mass. The surface temperature $T_s$ (red dashed curve) is set to the radiative equilibrium temperature. This is well-justified because the insolation is two orders of magnitude greater than the sensible and latent heat flux associated with evaporation as demonstrated by \citet{Kang-Ding-Wordsworth-et-al-2021:escaping} and \citet{Nguyen-Cowan-Pierrehumbert-et-al-2022:impact}. In the limit $R_{\star}+a \ll d$, $T_s$ follows 
\begin{equation}
    T_s=T_{s0}(\cos\theta_{TL})^{1/4},
    \label{eq:Ts}
\end{equation}
where $\theta_{TL}$ is the angular distance from the substellar point. However, the lava worlds considered here are typically very close to its host star, so close that the assumption $R_{\star}+a \ll d$ may not hold. An accurate surface temperature profile is crucial for TPW because it determines the location of magma ocean edge and mass deposition. We therefore adopt a more accurate $T_s$ formula inspired by \citet{Kopal-1954:photometric} (derivation available in the appendix). As can be seen in Fig.~\ref{fig:Ts-comparison}a, a planet with close-in orbit, which can be achieved at high $T_{s0}$ and low $M_\star$, would receive sunlight in the so-called ``permanent night-side''.

Near the substellar point, where $T_s$ is the highest, surface evaporation quickly fills the atmosphere with magma vapor made of sodium, SiO, Fe, Mg, Al$_2$O$_3$ etc. \citep[$F>0$, magenta curve, ][]{Miguel-Kaltenegger-Fegley-et-al-2011:compositions, Schaefer-Lodders-Fegley-2012:vaporization}, until a chemical equilibrium pressure $P_{\mathrm{chem}}$ is reached. Since the most volatile species, sodium, only accounts for $0.29\%$ of the mass assuming a Bulk Silicate Earth composition \citep{Schaefer-Lodders-Fegley-2012:vaporization}, we assume that it has been exhausted by atmospheric escape \citep{Kang-Ding-Wordsworth-et-al-2021:escaping} or by mass transport to the night-side \citep{Kite-Fegley-Schaefer-et-al-2016:atmosphere}, and that the atmosphere is dominated by SiO.
On the night-side, the chemical equilibrium pressure is much lower due to the low surface temperature, and that creates a pressure gradient (shown by black curve), which accelerates the gas from zero to O($10^3$)m/s (blue curve). The air temperature $T_a$ (red solid curve) does not necessarily remain close to the local $T_s$ because of the heat advection from upstream and because of the inter-conversion between internal energy and the flow's kinetic energy by pressure work.
After the atmosphere travels a certain distance away from the substellar point, the surface temperature and the chemical-equilibrium pressure drop so much that the atmosphere starts to be absorbed back into the magma ocean ($F<0$, magenta curve). Further away, beyond the magma ocean edge, the surface consolidates and the chemical equilibrium pressure drops to almost zero. However, the mass flux remaining in the atmosphere $\Phi$ (green curve) will not decay to zero immediately because the mass exchange $F$ between the magma ocean/lithosphere and the atmosphere is not infinitely efficient. Instead, $F$ is limited by the molecular collision frequency at the interface. Following \citep{Ingersoll-Summers-Schlipf-1985:supersonic, Kang-Ding-Wordsworth-et-al-2021:escaping}, the mass exchange rate $F$ can be estimated as
\begin{eqnarray}
   \label{eq:F-P}
  F&=&\begin{cases}\frac{\alpha P_a}{\sqrt{2\pi RT_a}}\left(\frac{P_{\mathrm{chem}}(T_s)}{P_a}-1\right) & \mathrm{within~magma~ocean}\\
    -\frac{\alpha P_a}{\sqrt{2\pi RT_a}} & \mathrm{out~of~magma~ocean}\end{cases}
\end{eqnarray}
where $T_a,\ P_a$ are the atmosphere temperature and pressure, $T_s$ is the surface temperature, $P_{\mathrm{chem}}$ is the pressure required by chemical equilibrium in the magma ocean, $R$ is gas constant, and $\alpha$ is the collision efficiency. Here we assume $\alpha=1$, i.e., each collision results in a molecule exchange to reduce disequilibrium, and we ignore the fact that the exchange efficiency is usually less than one, especially when the surface temperature is relatively warm \citep{Haynes-Tro-George-2002:condensation}.

As the magma ocean is composed of liquid, we make the assumption that the sea surface height will decrease uniformly everywhere to ensure that the surface remains near the geoid\footnote{Having an atmospheric flow over 1000~m/s blowing above the magma ocean, we expect the sea surface height to be slightly decreased near the substellar point and increased near the magma ocean edge. This will further enhance the negative tidal bulge induced by mass redistribution $L_a$ and make the system more prone to true polar wander.}. To account for the mass redistribution within the magma ocean, we calculate the average mass loss rate within the magma ocean and assign that value to everywhere inside the magma ocean. This modified mass deposition rate $\tilde{F}$ is shown in Fig.~\ref{fig:example-case-mass-deposition-load-width}b for cases with various surface temperatures and planetary masses. Solid curves show the mass accumulation rate out of the magma ocean as a function of angular distance from the substellar point and the height of the dots shows the total mass loss rate inside the magma ocean. There are two notable trends that we want to highlight. Both are fairly straightforward. First, as the surface temperature increases, chemical equilibrium pressure $P_\mathrm{chem}$ increases and that enhances atmospheric transport and mass deposition. Secondly, as the planet becomes more massive (denoted by thicker lines and bigger dots), the gravity strengthens, so that the same surface pressure will correspond to a lower column mass in the atmosphere. This will then lead to a decrease of mass deposition rate.

Results shown here assume $M_\star=10\times$ solar mass. Decreasing stellar mass does not change the results here to any substantial extent, except that mass deposition tends to occur closer to the anti-stellar point. This is because, as stellar radiation dims, the same $T_{s0}$ can only be achieved at much more close-in orbits, and that would allow stellar radiation to expand toward the night-side (see Fig.~\ref{fig:Ts-comparison}a). 

  \subsection{Equilibrium mass load.}

  
 While atmospheric transport continuously builds up the mass load, the silicate mantle beneath will tend to relax, allowing an equilibrium state to be reached. To estimate the equilibrium mass load, we thus need to estimate the mantle relaxation timescale. The relaxation timescale varies with the spatial scale of the load $l^{L}$. To get an estimate, we find out the e-fold decay scale of the $F$ profile outside the magma ocean and that compares reasonably well with $\pi R T_{s0}/g$ as shown by Fig.~\ref{fig:example-case-mass-deposition-load-width}c ($g$ is gravity). The scaling $\pi R T_{s0}/g$ is the inverse decay rate obtained by balancing the two sides of the mass continuity equation $d(P_aV_a\sin\theta_{TL}/g)/(a~ d\theta_{TL})=F\equiv -P_a/\sqrt{2\pi RT_a}$ assuming $V_a\sim \sqrt{2\pi RT_a}$ and $T_a\sim T_{s0}/2$.

  As shown in Fig.~\ref{fig:example-case-mass-deposition-load-width}c, the load wavelength is in order of several hundred to several thousand kilometers, which is likely to be greater than the lithosphere thickness\footnote{The lithospheric thicknesses of Earth and Mars are of order 100~km and 300~km, respectively; super-Earths are expected to have thinner lithospheres in order to permit escape of their greater internal heat generation}, implying that relaxation will be controlled primarily by the mantle beneath. This situation is analogous to post-glacial rebound on Earth \citep{Cathles-1975:viscosity}. For an isoviscous convecting mantle the relaxation timescale can be estimated by
  \begin{equation}
    \label{eq:tau-rlx}
    \tau_{\mathrm{rlx}}=\frac{2\eta_{\mathrm{man}} k^{L}}{\rho_{\mathrm{man}}g},
  \end{equation}
  where $\eta_{\mathrm{man}}$ and $\rho_{\mathrm{man}}$ are the viscosity and density of the mantle and $k^{L}\equiv 2\pi/l^{L}$ is the wavenumber of the load, ignoring the effect of the thin elastic lithosphere on the top. This yields a relaxation timescale of 30-100~kyr as summarized in Table.~\ref{tab:relaxation-timescale}. The support by lithosphere will become increasingly important as the planetary size decreases, which would allow the load to be maintained for longer period of time. Thus, it is likely that $\tau_{\mathrm{rlx}}$ and thus the chance for TPW is underestimated for small planets. Note that the relaxation timescale is at least 2-3 orders of magnitude shorter than the lifespan of the star we consider here (10 solar mass star has a lifespan of 20 Myr).

\begin{table*}[hptb!]
  \begin{tabular}{ccccccccc}
    \hline
     ~~    &  0.01Me  &  0.02Me &   0.05Me  &  0.1Me &   0.2Me  &  0.5Me &   1Me &   3Me\\
    \hline
    2000K  &   101   &     93   &     80   &    70   &    60   &    49   &   40  &   27 \\
    3000K  &    60   &     63   &     57   &    52   &    44   &    38   &   34  &   29 \\
    4000K  &    48   &     45   &     45   &    42   &    39   &    35   &   33  &   29 \\
    \hline
     \end{tabular}
  \caption{Relaxation timescale $\tau_{\mathrm{rlx}}$ of the load induced by atmospheric transport for a selective set of substellar temperatures (row) and planetary masses (column). $M_\star$ is set to $10\times$solar. Units: kyr. }
  \label{tab:relaxation-timescale}
\end{table*}

When a balance is reached between the addition of a load by atmospheric transport on the one hand and the relaxation of the load due to viscous mantle flow on the other, the equilibrium mass load
\begin{equation}
    M^{L}=\tilde{F}\tau_{\mathrm{rlx}}, \label{eq:mass-load}
\end{equation}
where $\tilde{F}$ is the modified mass deposition rate shown in Fig.\ref{fig:example-case-mass-deposition-load-width}b.
Dividing $M^{L}$ by $\rho_{\mathrm{litho}}$, we obtain the equilibrium height of the load $h^{L}$, which is shown by Fig.~\ref{fig:heq-Qload}a, increases with $T_{s0}$ and decreases with $M_p$ just like the mass deposition rate $F$.

\subsection{True polar wander criterion.}
In order for true polar wander (TPW) to occur, the equilibrium mass load needs to dominate the permanent component of the tidal bulge. By calculating the influence of the rotational tidal bulge and the mass load on the orientations of the principal axes of the inertial tensor, \citet{Matsuyama-Nimmo-2007:rotational} and \citet{Nimmo-Matsuyama-2007:reorientation} derived the following criterion for TPW:
\begin{equation}
  \label{eq:Q-criterion}
  Q^L\equiv\frac{G M_p}{\Omega^{2}a^{3}} \frac{3 \sqrt{5} G_{20}^L}{\left(k_{f}^{T*} -k_{f}^{T}\right)}>1,
\end{equation}
where 
$G_{20}^L$ is the normalized $Y_{20}$ mode of the equilibrium mass load $M^L$, $k_{f}^{T*}$ and $k_{f}^{T} $ are the long-term degree-2 Love numbers of the planet without and with an elastic layer, respectively. Here, $k_f^{T*}-k_f^T$ is computed using the two-layer model formula given by \citet{Harrison-1963:analysis} and represents the fraction of the tidal bulge which is permanent. To be conservative we assume the whole mantle is elastic, which maximizes $k_f^{T*}-k_f^T$ (always of order unity) and minimizes $Q$. The quantity $Q^L$ is dimensionless as can be easily verified. Only the $Y_{20}$ mode appears in the above formula because the definition of inertia tensor $\delta I_{i j}=\int_{V} d V \delta \rho\left(r^{2} \delta_{i j}-r_{i} r _j\right)$ involves a projection of density anomaly $\delta \rho$ onto a factor $\left(r^{2} \delta_{i j}-r_{i} r _j\right)$, whose angular structure resembles that of $Y_{20}$.

Our mass load $M^L$ is symmetric about the tidal axis instead of the poles, so to compute $G_{20}^L$, we need to use the addition theorem to rotate the poles. This finally leads us to
\begin{equation}
  \label{eq:G20}
  G_{20}^L=\frac{\pi a^2}{5M_p}\int_0^\pi M^L(\theta_{TL})Y_{20}(\theta_{TL})\sin\theta_{TL}d\theta_{TL},
\end{equation}
where $Y_{20}$ is normalized such that $\iint_0^\pi 2\pi Y_{20}Y_{20}\sin\theta_{TL}d\theta_{TL}=4\pi$.

\begin{figure*}[htbp!]
    \centering
    \includegraphics[width=\textwidth]{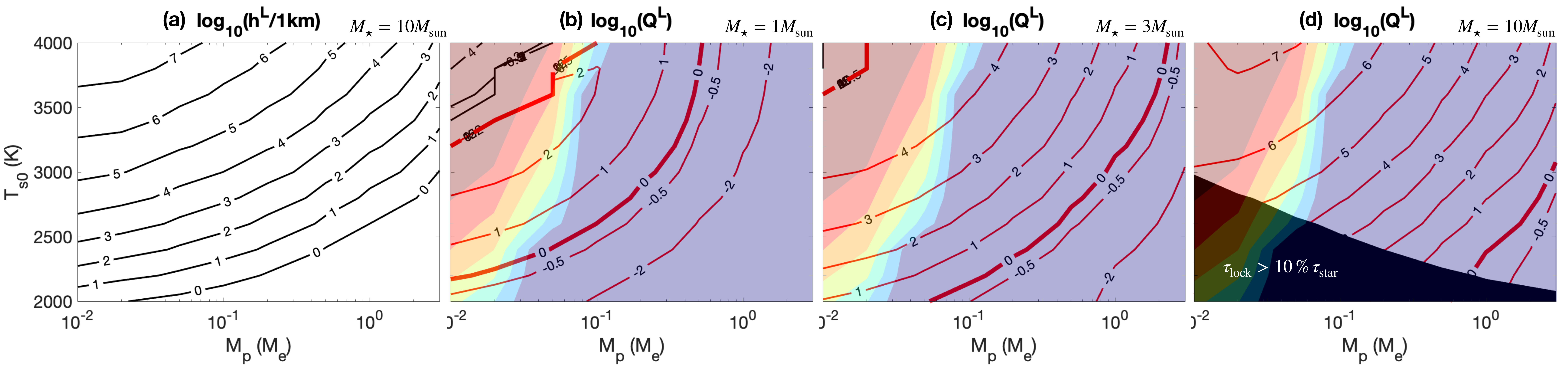}
    \caption{\small{(a) shows the equilibrium height of the load (assuming $M_\star=10\times$solar mass, other solar masses yield similar results). (b-d) show the nondimensional load size $Q$ as a function of insolation level (measured by the substellar point temperature $T_{s0}$) and planetary mass $M_p$, for $M_\star=$ 1, 3 and 10 $M_{\mathrm{sun}}$, respectively. Red contours denote positive $Q^L$ and black contours denote negative $Q^L$. The black shading in (b-d) marks the parameter combinations where the timescale for a specific planet to get tidally-locked $\tau_{\mathrm{lock}}$ is greater than 10\% of the lifespan of the star $\tau_{\mathrm{star}}$. The colorful shading overlaid shows atmospheric escape rate obtained from the analytical model in \citet{Kang-Ding-Wordsworth-et-al-2021:escaping}. Red color corresponds to escape rate beyond $10^{12}$~kg/s and blue color corresponds to escape rate below $10^2$~kg/s. With an escape rate of $10^7$~kg/s, it will take 0.2Gyr to completely vaporize a planet of 0.01 earth mass. In panels b-d, areas encompassed by the bold 0-contour are likely to undergo TPW. }}
    \label{fig:heq-Qload}
  \end{figure*}

The final results of $Q^L$ is shown in Fig.~\ref{fig:heq-Qload}b-d for a host star with $1\times$, $3\times$ and $10\times$ solar mass, respectively. The thick 0-contour in each figure demarcates the regime where TPW can occur from the regime where it cannot. For a given star mass, TPW tends to happen on planets with higher temperature and lower mass due to the stronger atmospheric mass transport, longer relaxation timescale and the smaller rotational tidal bulge. It is worth noting that, with $M_\star=1\times$ solar mass, the $Q^L$ for planets with super small size and high surface temperature start to become negative. This is because as $(a+R_\star)/d$ increases, stellar radiation starts to reach the so-called permanent night-side, pushing the magma ocean edge and deposition zone toward the anti-stellar point (see Fig.~\ref{fig:Ts-comparison}a). Depositing mass near anti-stellar point would stabilize rather than destabilize the orientation of the planet. However, for these small hot planets, atmospheric escape may be quite significant and our transport model may not be applicable. To quantify this, we show the analytical approximation for atmospheric escape rate \citep[see section~5 of][]{Kang-Ding-Wordsworth-et-al-2021:escaping} in shadings, with reddish colors denoting significant escape ($>10^{12}$~kg/s) and blueish colors denoting negligible escape ($10^2$~kg/s).
s
Comparing between different panels, it can be easily seen that planets are more prone to TPW when the host star is massive. This is because the star's luminosity increases with the cubic power of the star mass \citep{Phillips-2013:physics}, and having a high luminosity necessarily means a planet receiving a specific amount of insolation is farther away from the star. At a farther distance, the planet's rotational and tidal bulges will both decrease, making them easier to be dominated by the load. A greater star mass, however, brings about a second consequence -- a shorter life span of the star. If the star's lifespan is too short, the planet may not have enough time to become tidally locked, which is assumed by the atmospheric transport model (Step-1). In Fig.~\ref{fig:heq-Qload}b-d, we mark the parameter space with $\tau_{\mathrm{lock}}$ greater than 10\% of the lifespan of the star $\tau_{s}$ (see Table.\ref{tab:parameters}) by black shading. Here the timescale for tidal locking is estimated by
\begin{equation}
\label{eq:taulock}
    \tau_{\mathrm{lock}}=\frac{2\Omega d^{6}M_p Q_f}{15 G M_{\star}^2 k_{2} a^{3}},
\end{equation}
where $k_2$ is the degree-2 tidal Love number of the planet, which we estimate using the formula by \citet{Harrison-1963:analysis}, and $Q_f$ is the quality factor for the planet, which is set to 100.

  \section{Concluding remarks.}
  Our main finding is that lava planets within a certain mass and temperature range (the $\log_{10}(Q^L)>0$ area in Fig.~\ref{fig:heq-Qload}b-d) may undergo true polar wander (TPW), and this range widens as star mass $M_\star$ increases as can be seen by comparing the b,c,d panels in Fig.~\ref{fig:heq-Qload}. Once the TPW is initiated, the planet's static tidal bulge and the location of mass deposition will adjust accordingly to be aligned with the new tidal axis. These adjustments would allow the planet to continuously turn at a certain rate. As proposed by \citet{Kite-Fegley-Schaefer-et-al-2016:atmosphere}, a lava planet's atmosphere may become more and more enriched in refractory material as volatile species are transported to the night-side. Over geological timescales, the major composition of the atmosphere may evolve from being Na, K dominant, to Mg, Si, Fe dominant, and eventually to Al, Ca dominant. If TPW happens, such evolution of atmospheric composition may stop because the volatile species piled up over the solid mantle can now be melted and revaporized as they are reoriented to near the substellar point. The correlation between the likelihood for TPW and the star mass indicates that lava planets surrounding more massive stars may have a less refractory-enriched ocean and atmosphere. The shorter lifespan of a massive star would also result in a less evolved atmospheric and magma composition. This effect must be taken into account to isolate the differences induced by TPW.

 Besides the atmosphere/mantle composition, we also expect ultra-hot rocky planets to be aspherical. Not only will mass build up out of the magma ocean, but also the momentum torque induced by the supersonic atmospheric flow, the gradient of atmosphere-induced normal pressure and the mass redistribution inside the magma ocean (evaporation near substellar point and condensation near the magma ocean edge) all tend to flatten the planet along the tidal axis. Since the timescale for TPW is likely in the order of 10-100My \citep{Tsai-Stevenson-2007:theoretical}, much longer than the relaxation timescale $\tau_{\mathrm{rlx}}$ (Table.\ref{tab:relaxation-timescale}), the equilibrium surface topography, therefore, should be controlled by $\tau_{\mathrm{rlx}}$. As shown by Fig.\ref{fig:heq-Qload}a, $h^L$ can be comparable to the size of the planet for some cases! This may also be detected in the future.

 Finally, we want to highlight some assumptions that have been made in this work. First, we assume that the refractory material left behind after evaporation will be advected to the magma ocean edge and accumulate there under the surface stress exerted by the supersonic atmospheric flow. We get preliminary results from a lava ocean circulation model that is in line with this assumption. This way, the refractory material (Ca, Al minerals) will not significantly alter the surface albedo or the composition of the magma ocean, or inhibit the mass exchange efficiency between the atmosphere and the ocean ($\alpha$), before the mass redistribution required for TPW is achieved. Second, in the atmospheric transport model, we ignore the rotation effect on the air flow, which, if accounted for, would induce flow asymmetry between the leading and trailing hemispheres and thereby a preferred direction for TPW to occur. The importance of rotation effect can be measured by the smallness of the Rossby number, Ro$=U/(\Omega L)\sim \sqrt{RT_{s0}}/(\Omega a)$. The smallest Ro is achieved at the lowest star mass, highest planet mass and highest surface temperature, and that combination (3 earth-mass planet surrounding 1 solar mass star with 4000~K substellar temperature) yields a Rossby number of $0.5$, which indicates only a moderate rotational effect. For planets surrounding a 3x solar mass star, the smallest Rossby number we get is $3$, and for planets around a 10x solar mass star, the smallest Rossby number is $27$. These all point to rotation playing a secondary role in the atmospheric transport. More work using 2D or 3D transport models are needed to quantify this effect.

\begin{acknowledgments}
   WK acknowledge supports from her startup funding from the department. We appreciate insightful comments and communications with Prof.~Raymond Pierrehumbert at Oxford and Prof.~Edwin Kite at University of Chicago.
\end{acknowledgments}

%





\appendix
\renewcommand{\thefigure}{A\arabic{figure}}

\setcounter{figure}{0}

\section{Distribution of illumination on a close-in planet}
\label{sec:illumination-close-in}

   \begin{figure*}[htbp!]
    \centering
    \includegraphics[width=0.8\textwidth]{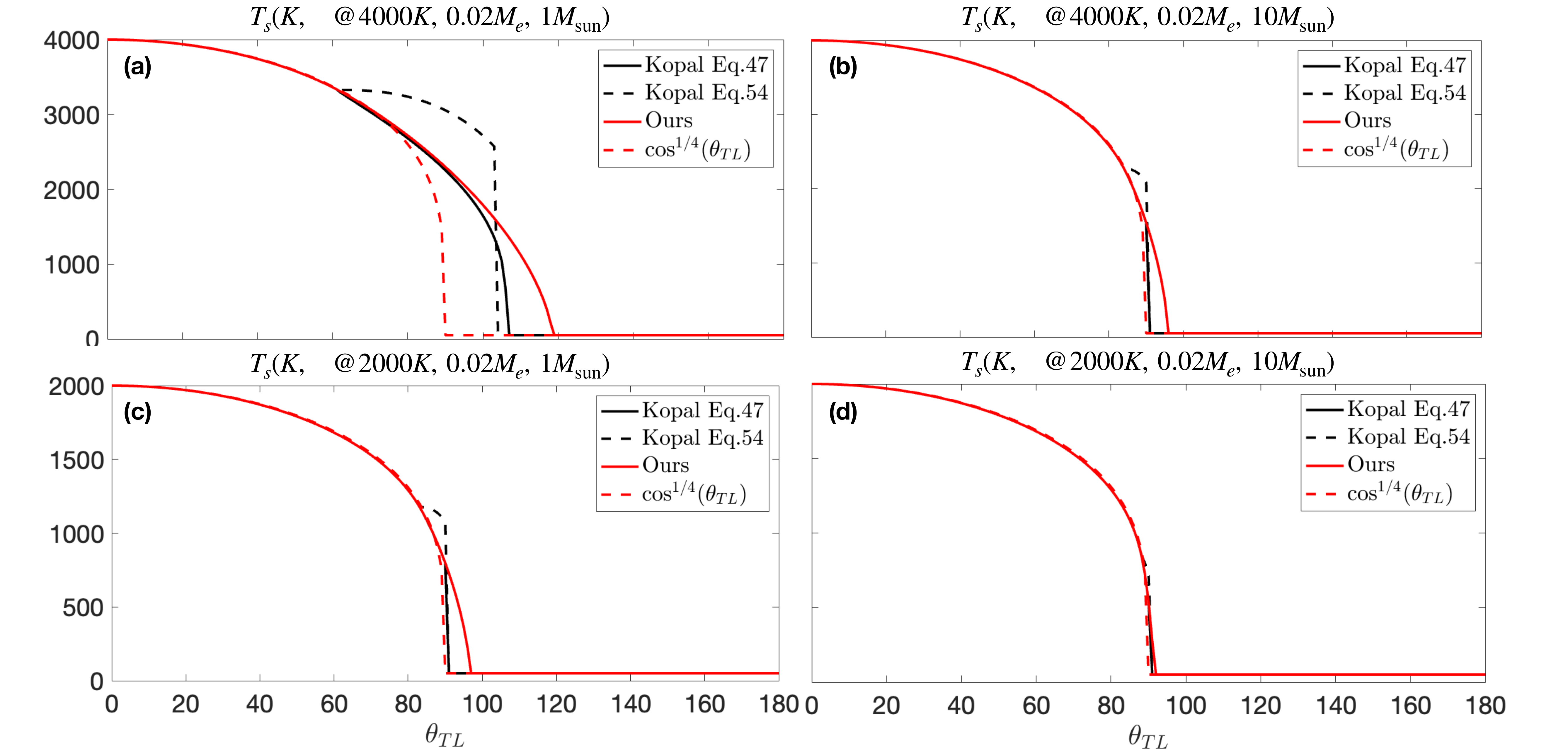}
    \caption{\small{Radiative equilibrium surface temperature profiles assuming various substellar temperatures and stellar masses.}}
    \label{fig:Ts-comparison}
  \end{figure*}
  
\begin{figure*}[htbp!]
    \centering
    \includegraphics[width=0.8\textwidth]{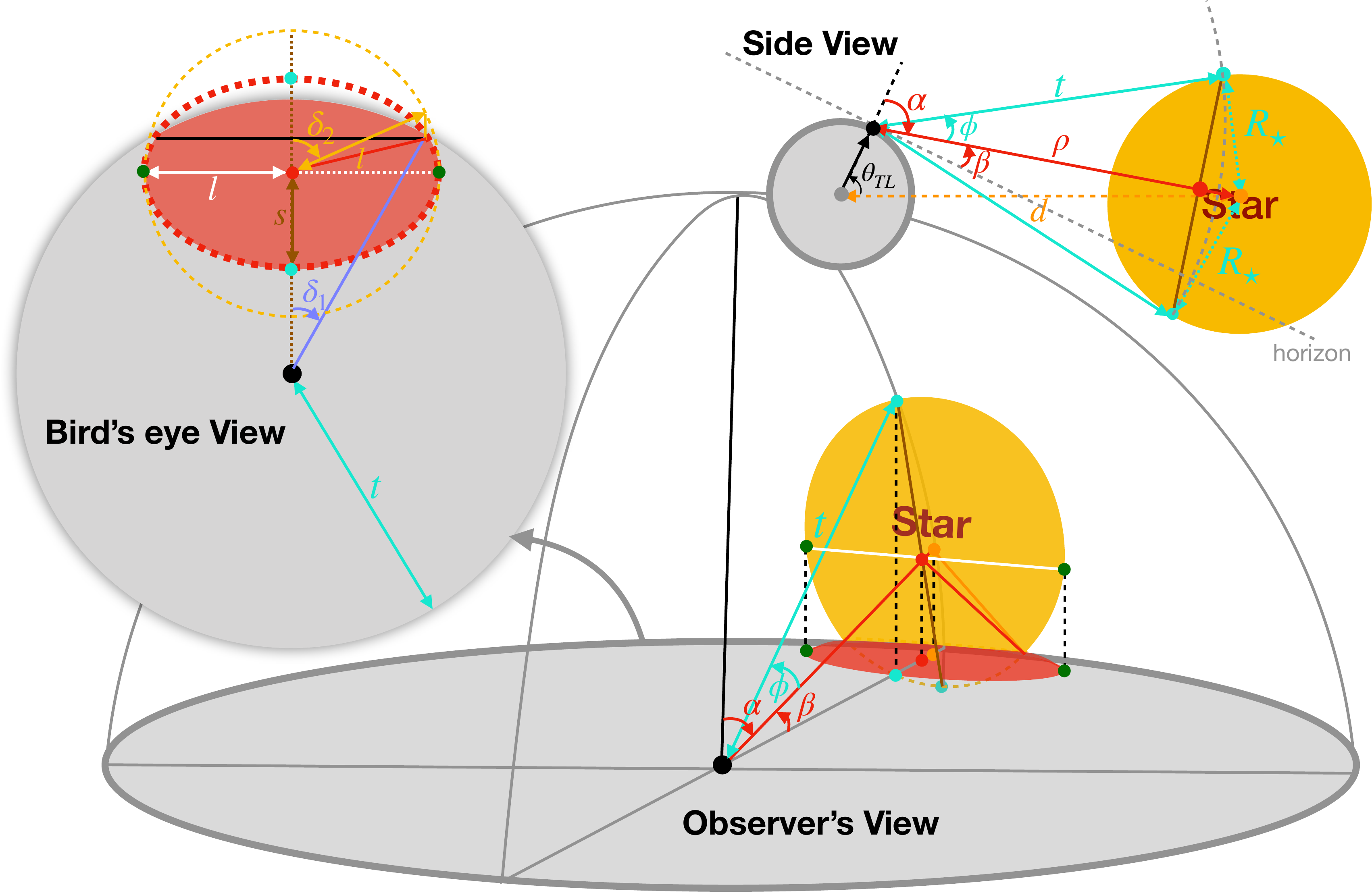}
    \caption{\small{Geometry from the point of view of an observer at the surface of the planet. The side view demonstrates the geometric relationship between the planet (grey) and the star (orange). The bird's eye view demonstrates the shadow (red shading) cast by stellar disk (orange shading) on the horizon plane (grey shading). The red circle is the projected stellar disk, with key intersections marked in the same color as the side-view schematics. Orange circle is the unprojected stellar disk (orange shading in side view) overlaid rotated to the horizontal plane. See text for more details. }}
    \label{fig:geometry}
  \end{figure*}

As the distance between the host star and the planet decreases, the sun appears bigger in the sky and the terminator broadens up. This means an increasing portion of the planet will be illuminated by part of the star (permanent sunset/sunrise), and a decreasing portion of the planet sees the complete host star or none. The fully illuminated co-latitudes lie in
\begin{equation}
  \label{eq:theta-fully-illuminate}
  \theta_{TL}<\theta_{f}\equiv\arccos((a+R_\star)/d),
\end{equation}
whereas those in complete darkness lie in
\begin{equation}
  \label{eq:theta-fully-illuminate}
  \theta_{TL}>\theta_{d}\equiv\arccos((a-R_\star)/d),
\end{equation}
within which $0<\theta_f,\ \theta_d<\pi$.
This photometric effect could potentially important to true polar wander (TPW) because it can affect the location of the magma ocean edge and hence atmospheric deposition, and thereby, the gravitational torque induced. This matter has been considered when people tried to estimate the reflection in close binary systems \citep{Kopal-1954:photometric} and is applied in recent work on lava worlds \citep{Nguyen-Cowan-Banerjee-et-al-2020:modelling, Nguyen-Cowan-Pierrehumbert-et-al-2022:impact}. The radiation distribution $I(\theta_{TL})$ within the fully illuminated region ($\theta_{TL}<\theta_{f}$) can be written down analytically
\begin{equation}
  \label{eq:rad-fully-illuminate}
  F(\theta_{TL})=\frac{L_\star}{4\pi}\frac{d\cos\theta_{TL}-a}{\rho^3}
\end{equation}
where $\rho\equiv \sqrt{d^2+a^2-2da \cos\theta_{TL}}$ is the distance between the observer and the center of the star (see Fig.~\ref{fig:geometry} side view). When trying to evaluate the radiation between $\theta_f$ and $\theta_d$, \citep{Kopal-1954:photometric} only retains the lowest order terms associated with the photometric effect (see their Eq.42 for assumptions), which leads to the formula given by their Eq.~47. Then further approximation is made to simplify this formula to Eq.54, where radiation $F$ is expressed only as a function of the zenith angle of the center of the star $\alpha$. This radiation can then be used to evaluate the radiative equilibrium surface temperature
\begin{equation}
  \label{eq:Ts-illumination}
  T_s=(F/\sigma)^{1/4}
\end{equation}
Due to these simplifications, the estimated $T_s$ exhibits an unphysical bump before it finally decays to zero as $\theta_{TL}\rightarrow \theta_d$ as can be seen from Fig.~\ref{fig:Ts-comparison}.
  
To better deal with this, we decide to reevaluate the radiation received in this penumbra zone. Assuming that the stellar radiation is isotropic (ignoring edge dimming effect), the radiation received at $\theta_{LT}$ only depends on the solid angle of the stellar disk projected onto the horizon plane. In the planet's sky, the star always appears like a perfect circle unless it is underneath the horizon. Fig. \ref{fig:geometry} shows a schematics of the geometries. At a given tidally-locked co-latitude $\theta_{LT}$, the angle between the center of the stellar disk (orange shading) and the norm of the local planetary surface (zenith angle corresponding to the center of the star) is
\begin{equation}
  \label{eq:zenith-angle}
  \alpha=\arccos((d\cos\theta_{TL}-a)/\rho),
\end{equation}
and the angular span of the star is
\begin{equation}
  \label{eq:span-angle}
  \phi=\arcsin(R_\star/\rho).
\end{equation}
The isotropic approximation guarantees that the radiance is a constant over the entire stellar disk. The total radiation per area reaches $\theta_{LT}$ is proportional to the proportion of the horizon plane ``shadowed'' by the stellar disk.
In the bird view (see Fig.\ref{fig:geometry}), we can see the shadowed area should be elliptical, with semi-major axis $l$ and semi-minor axis $s$ equal to
\begin{equation}
  \label{eq:ellipse}
  l=t\sin\phi,\ s=t\sin\phi\cos\alpha,
\end{equation}
where $t=\rho\cos\phi$ is the radius of the horizon circle. In the fully illuminated zone ($\theta_{LT}<\theta_f$), the elliptical shadow is completely inside the horizon circle, so that the percentage of horizon plane shadowed by the stellar disk can be written as
\begin{equation}
  \label{eq:shadow-area-fully-illuminate}
  J_f=\frac{\pi ls}{\pi t^2}=\sin^2\phi\cos\alpha=\frac{(d\cos\theta_{TL}-a)R_\star^2}{\rho^3}.
\end{equation}
The radiation and surface temperature at $\theta_{LT}$ ($<\theta_f$) can then be written as
\begin{equation}
  \label{eq:insolation-Ts--fully-illuminate}
  F(\theta_{TL}<\theta_f)=\sigma T_\star^4J_f, T_s(\theta_{TL}<\theta_f)=T_\star\left(\frac{(d\cos\theta_{TL}-a)R_\star^2}{\rho^3}\right)^{1/4}.
\end{equation}
This matches the results given by \cite{Kopal-1954:photometric}, which is repeated in Eq.\ref{eq:rad-fully-illuminate}. At substellar point, the equilibrium surface temperature equals $T_{s0}\equiv T_s(0)=T_\star\sqrt{\frac{R_\star}{d-a}}$. This is used to compute the distance $d$ between the planet and its host star in order to achieve a specific substellar temperature $T_{s0}$.

In the penumbra zone, the elliptical shadow will intersect with the edge of the horizon circle. The percentage of horizon plane shadowed by the stellar disk equals
\begin{equation}
  \label{eq:shadow-area-penumbra}
  J_p=\frac{ls}{\pi t^2}\left[\pi-\delta_2+\sin\delta_2\cos\delta_2\right]+\frac{t^2}{\pi t^2}\left[\delta_1-\sin\delta_1\cos\delta_1\right],
\end{equation}
where
\begin{eqnarray}
  \delta_1&=&\arccos\left(\cos\phi/\cos\beta \right) \label{eq:delta1}\\
  \delta_2&=&\arccos\left(\tan\beta/\tan\phi\right)\label{eq:delta2},
\end{eqnarray}
where $\beta=\pi/2-\alpha$. When $\beta<0$ (i.e. $\alpha>\pi/2$), the center of stellar disk is below the horizon, therefore, $\pi/2<\delta_2<\pi$.

This then leads to the following equilibrium surface temperature,
\begin{equation}
  \label{eq:Ts-fully-illuminate}
  T_s(\theta_f<\theta_{TL}<\theta_b)=T_\star\left(\frac{\sin^2\phi\cos\alpha}{\pi}\left[\pi-\delta_2+\sin\delta_2\cos\delta_2\right] + \frac{1}{\pi}\left[\delta_1-\sin\delta_1\cos\delta_1\right]\right)^{1/4}.
\end{equation}

The above results are shown for four cases with various substellar temperature $T_{s0}$ and stellar mass $M_\star$ in Fig.~\ref{fig:Ts-comparison}, compared against the traditional $\cos^{1/4}\theta_{TL}$ profile, and the two approximated formula given by \cite{Kopal-1954:photometric}.
  
\bibliography{export}{}
\bibliographystyle{aasjournal}



\end{document}